%% file: Pixel2002.tex
\documentclass[12pt]{article}
\usepackage{epsfig}

\input econfmacros.tex
\def\Title#1{\begin{center} {\Large {\bf #1} } \end{center}}

\begin{document}

\Title{Test Beam Results of ATLAS Pixel Sensors }

\bigskip\bigskip


\begin{center}  

\begin{large}  
Tommaso Lari\index{Lari, T.}
\end{large}

{\it INFN, Sezione di Milano, Via Celoria 16, I-20133 Milano, ITALY}
\bigskip\bigskip

On behalf of the ATLAS Pixel Collaboration

\end{center}

\begin{center}  
{\bf Abstract}
\end{center}

Silicon pixel detectors produced according to the ATLAS Pixel Detector design 
were tested in a beam at CERN in the framework of the ATLAS collaboration.
The detectors used n$^+$/n sensors with oxygenated silicon substrates. The 
experimental behaviour of the detectors after irradiation to 
$1.1 \cdot 10^{15}$~\nequ and 600~kGy is discussed. At the sensor bias 
voltage of 600~V the depleted depth is measured to be 229~$\mu$m, 
the mean collected charge is 20000 electrons, the detection efficiency 
is 98.2\% and the spatial resolution is 9.6~$\mu$m.

\section{Introduction}

At the LHC silicon microstrip and pixel detectors will be exposed to 
unprecedented levels of ionizing and hadronic radiation. As a consequence 
of displacement damage arising from hadronic radiation, the silicon detectors
will change their characteristics during their 
lifetime~\cite{Lindstrom:ww,Li02}. 
Potential radiation-induced performance degradation 
is a serious issue, which has been studied in detail.

%
%
%
%
%
%

The ATLAS Pixel Detector~\cite{PDTDR,Gar02} is required to be tolerant to 
an hadronic  fluence~\footnote{In this paper the 
fluences are converted in \nequ where $n_{\mbox{eq}}$ is the number of 
particles with the non-ionizing energy loss of a 1~MeV neutron.}
 of $10^{15}$~\nequ and an ionizing dose of 550~kGy.
In order to verify their ability to meet these requirements,  
some pixel detectors produced according to the ATLAS design 
have been exposed to 110\% of the target fluence and dose and have been 
operated in a test-beam experiment at CERN. Measurements of the 
thickness of the depleted region, of the efficiency and uniformity of 
charge collection, of hit detection efficiency and spatial resolution 
have been performed and are reported here.

\section{Tested devices and experimental setup}  

The ATLAS Pixel sensors~\cite{Alam:1999yh}
consist of $\mbox{n}^+$ implants on a high 
resistivity n-bulk substrate. 
This choice allows the detector to be operated in partial depletion mode 
after type inversion due to irradiation, 
as the n-p junction and the depleted zone are located 
near the pixels. The substrate is oxygenated to improve its radiation 
hardness.

The 
pixel dimensions are $50 \times 400 \;\mu\mbox{m}^2$ and the sensor thickness
is 250~$\mu$m. The isolation between pixels is achieved with the 
p-{\it spray} technique~\cite{Alam:1999yh,Richter:fs}. A network of 
bus lines, one every other column pair of pixels, is connected to an 
$\mbox{n}^+$ implant dot near each pixel. This {\it bias grid} allows to 
bias the detector during testing, before attachment of the front-end 
electronics.

The irradiated devices had a $8 \times 8 \;\mbox{mm}^2$ sensor connected 
to a front-end electronics 
chip~\cite{Ein02,Bec02,Ric02} of the same size. In ATLAS, pixel sensors 
will have a size of $16 \times 64 \;\mbox{mm}^2$ and will be connected to 
16 front-end chips.

Every pixel cell in the sensor is connected via bump-bonding\cite{PDTDR}
to a matching cell in 
the front-end chip, which includes a charge preamplifier and a discriminator. 
The pulse height is obtained from the time the signal stays over the 
discriminator threshold. The Time-Over-Threshold (ToT) 
is calibrated~\cite{Ric02}
for each pixel by injecting a known charge, with 
an accuracy of about 10\%. During test-beam operation the pixel thresholds 
had a mean value of 3000 electrons with a dispersion of 300 electrons rms. 
The noise was 400 electrons.

Irradiation was performed after bump-bonding 
at the CERN PS with a beam of 20~GeV protons~\cite{Saa02}. 
The detectors received the total ionizing 
dose of 600~kGy, corresponding to a fluence of  $1.1 \times 10^{15}$~\nequ. 

The test beam was performed at the CERN SPS accelerator with a pion beam 
of 180 GeV/$c$ momentum. A beam telescope~\cite{Treis:2002tx} consisting of 
4 pairs of double-sided silicon microstrip modules was used to track the 
particles. The position resolution of tracks projected onto the 
devices under test was 6~$\mu$m. 

During irradiation, between irradiation and the test beam operation 
and during the test beam operation the devices were kept below 0~$^0$C, 
suppressing the 
time dependence of defect concentration in the silicon. At the LHC, 
the cumulative effect of defect annealing after 10 years of operation 
is predicted to reduce the initial concentration of defects, so that 
the results presented here are conservative.

\section{Thickness of depletion zone}

\begin{figure}[htb]
\begin{center}
\epsfig{file=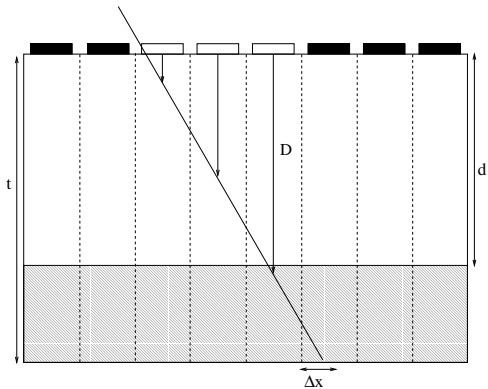,height=2in}
\caption{Schematic view of an irradiated sensor crossed by a track. The 
hatched zone corresponds to the not-depleted zone.}
\label{fig:dep1}
\end{center}
\end{figure}

The measurement of the depletion zone depth
is performed with  data taken 
with particle beam incident on the sensor at an angle of 30$^0$ 
respect to the normal to the pixel plane (Fig.~\ref{fig:dep1}), 
as described in~\cite{Gorelov:2001ca,Lari:2001ne}. 
A particle crossing the detector produces a cluster of hits consisting of 
the pixel cells that collect a significant fraction of the charge released 
in the subtended segment of the track. To measure the depth of the depleted 
region the maximum depth of track segment was used (Fig.~\ref{fig:dep1});
it was defined as the distance of the center of the track-segment subtended 
by a given pixel from the pixel itself. Since the entrance points of the  
tracks are uniformly distributed, all of the depths vary continuously and the 
maximum observed depth $D$ is a measurement of the depletion depth 
within the detector.

\begin{figure}[htb]
\begin{center}
\epsfig{file=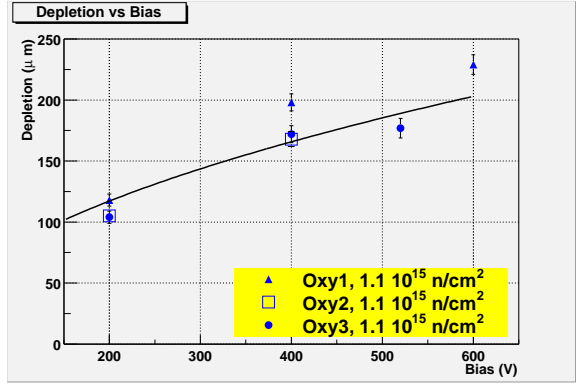,height=2in}
\caption{Measured depletion depth as a function of bias voltage, for three 
different irradiated assemblies. The solid curve is the model prediction.}
\label{fig:dep2}
\end{center}
\end{figure}

In Fig.~\ref{fig:dep2} the depletion depth measurements 
of the irradiated sensors
are plotted as a function of the applied bias voltage. The three different 
symbols refer to three sensors irradiated to the same fluence. The 
continuous line is the prediction based on the 
model~\cite{Lindstrom:ww} 
for the depletion depth of oxygenated sensors irradiated to 
$1.1 \times 10^{15}$~\nequ before annealing, assuming a uniform 
space charge, so that the depletion depth scales as the square root of the 
bias voltage. The measured depletion depth is in good agreement with 
this assumption. At 600~V, which is the maximum bias voltage foreseen 
in ATLAS, the sensors are almost fully depleted.

\section{Charge collection efficiency and uniformity}

The sum of the charges collected by the pixels in a cluster (cluster charge)
provides a measurement of the charge collection efficiency of the sensor. 

\begin{figure}[htb]
\begin{center}
\epsfig{file=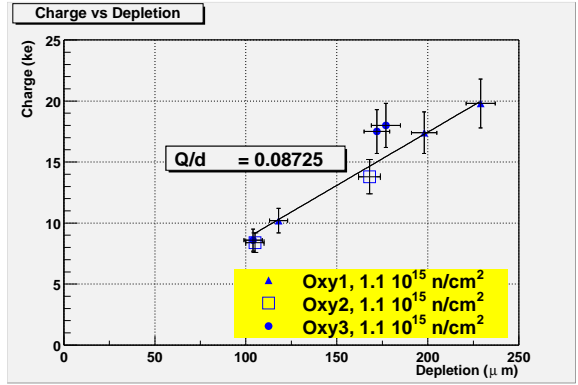,height=2in}
\caption{Average charge collected at normal incidence, as a function 
of depletion depth.}
\label{fig:cc1}
\end{center}
\end{figure}

In Fig.~\ref{fig:cc1} the average cluster charge at normal incidence 
is plotted 
against the depletion depth. As expected, the two quantities are proportional.
At 600~V an average signal of 20000 electron is seen, to be compared with 
25000 electrons collected by not irradiated and fully depleted 
sensors. Both numbers are 
affected by the 10\% uncertainty on the absolute scale provided by 
charge calibrations. 
The ratio between charge and depletion depth, coming from the fit to the 
data is $Q/d = 87 \pm 9 \;\mbox{e}^-/\mu$m.  

\begin{figure}[htb]
\begin{center}
\epsfig{file=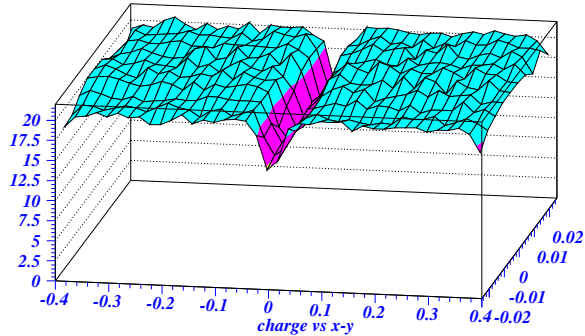,height=2in,angle=180}
\caption{Average charge collected (in thousands of electrons) 
as a function of the position inside a pair of pixel cells.}
\label{fig:cc2}
\end{center}
\end{figure}

The charge collection uniformity have been studied by considering the 
mean cluster charge as a function of the impact point of traversing particles
relative to the center of a pair of pixel cells 
(Fig.~\ref{fig:cc2}) as done in~\cite{Troncon:1999kw}
The coordinate axes $x,y$ are parallel to the short (50~$\mu$m) and 
long (400~$\mu$m) edges of the pixels, respectively, and are centered 
in $x$ at the centre of a single pixel and in $y$ between two adjacent pixels
belonging to a column pair. 

The uniformity is pretty good, except in the region between two columns 
where the bias grid is located ($y=0$). Here a significantly reduced signal 
is observed. No 
loss is observed between pixel columns where the bias grid is not 
present ($y = \pm 0.4$~mm) or between pixel rows.  

\section{Detection efficiency}

At the LHC the pixel detectors will be required to identify the bunch 
crossing which has produced a particle with an efficiency in excess of 97\%.
As the LHC bunch crossing rate is 
40~MHz, the charge collected by pixel must be detected
in a time window of 25~ns.  

Since the extracted beams used at CERN did not have an RF structure and the 
readout electronics was operated with a 40~MHz 
asynchronous clock (with respect to 
the particles), the phase between the arrival time of every particle and 
the clock was measured. During operation at the LHC this phase can be set 
for each individual module.
Further, for an accepted trigger of beam particle 
the pixel hits belonging to 16 contiguous ``bunch crossings'' (clock 
periods) were read out. This allowed to study the timing behaviour of the 
detectors~\cite{Ragusa:1999ky,And02}.

\begin{figure}[htb]
\begin{center}
\epsfig{file=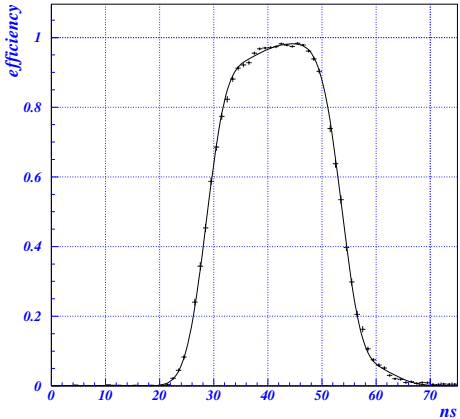,height=2.5in}
\caption{Detection efficiency versus time.}
\label{fig:eff}
\end{center}
\end{figure}

In Fig.~\ref{fig:eff} the detection efficiency is plotted against 
the difference between the beam trigger and the beginning of the clock 
period in which hits are looked for. The maximum efficiency is 98.2\% 
which is well within the Pixel Detector requirements. 

\section{Spatial resolution}

Spatial resolution is a key feature of the Pixel Detector, which strongly 
affects the vertex-finding performances of ATLAS. If there is no charge 
sharing between pixels, the hit position is reconstructed as the 
centre of the fired pixel, and the resolution $\sigma$ is related to the 
pixel pitch $p$ by $\sigma = p/\sqrt{12}$. If the charge released 
by the particle is shared by adjacent pixels, charge interpolation is 
possible, leading to improved resolution. The amount of charge sharing 
depends on the particle incidence angle, the thickness of depleted zone  
and the charge collection performances. 

In what follows, resolution along the short (50~$\mu$m) pixel direction 
will be considered.

\begin{figure}[htb]
\begin{center}
\epsfig{file=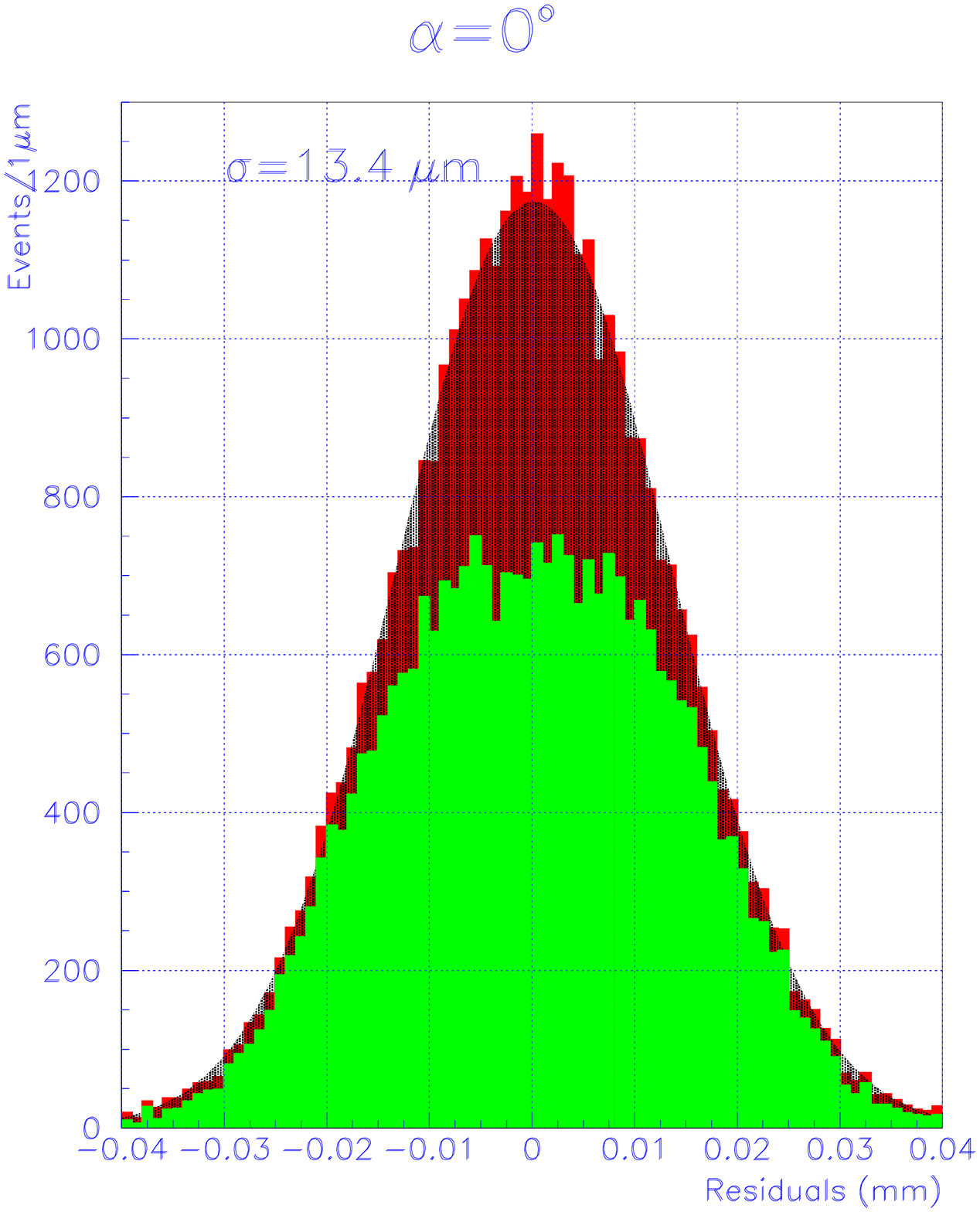,height=2in}
\epsfig{file=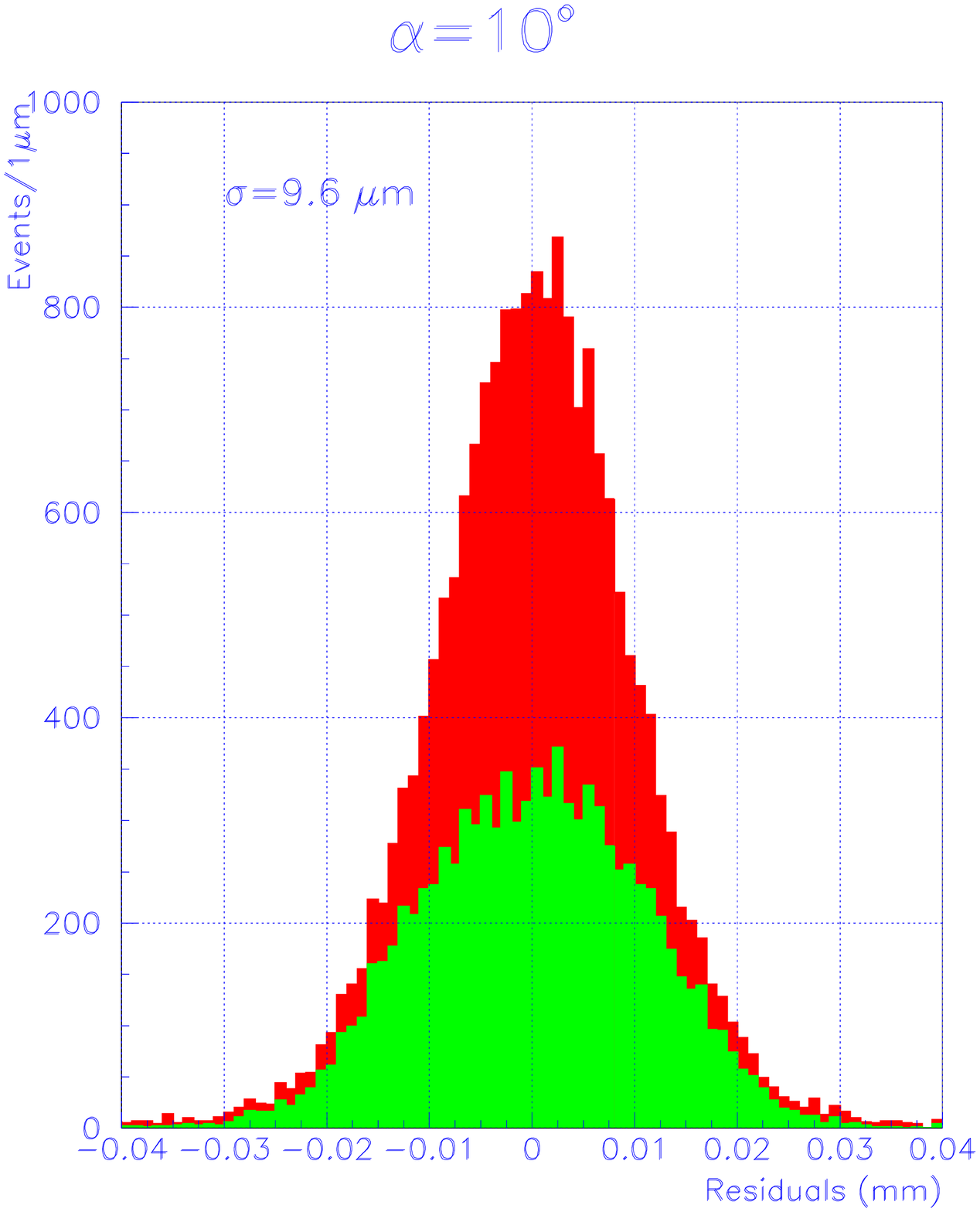,height=2in}
\caption{Residuals between the position measured by the pixel detector 
and by the beam telescope at normal incidence (left) and 10 degrees 
incidence angle (right). The lighter histogram 
is for one-pixel clusters, the darker one includes all clusters. }
\label{fig:res1}
\end{center}
\end{figure}

In Fig.~\ref{fig:res1} 
the distribution of the residuals between the 
hit position measured by the pixel detector and the track position 
provided by the beam telescope is shown for normal incidence and an 
incidence angle of $10^{0}$. The latter was chosen to be representative 
of ATLAS tracking conditions~\cite{Lar01}.
The lighter histograms 
represent the one-pixel cluster residuals, the darker ones include all 
clusters. The resolution is worse at normal 
incidence because of the larger fraction of one-pixel clusters, for 
which no charge interpolation is possible. A gaussian fit to the residuals 
distribution yields $\sigma = 13.4\;\mu$m at normal incidence and 
$\sigma = 9.6 \;\mu$m at $10^0$ incidence angle. 

\section{Conclusions}

ATLAS Pixel detectors with oxygenated silicon sensors and rad-hard 
electronics have been irradiated to $1.1 \times 10^{15}$~\nequ fluence 
and 600~kGy dose, 10\% more than the design values. They have been 
tested in a beam and were found to meet all ATLAS specifications.
The thickness of the depletion zone was found to be $229 \pm 8 \; \mu$m at 
600~V of reverse bias voltage. The mean signal height was 
$20000 \pm 10$\% electrons, to be compared with  $25000 \pm 10$\% electrons
before irradiation. The hit detection efficiency in  
a time window of 25~ns was 98.2\% and the spatial resolution at 10$^0$ 
incidence angle was 9.6~$\mu$m. 

\vspace{1cm}

{\bf \Large Acknowledgements}

The work reported here represents the joint effort of many individuals 
in the ATLAS Pixel collaboration. I would like to thank all of them, 
in particular John Richardson and Kevin Einsweiler from LBNL, Berkeley, and 
Attilio Andreazza, Francesco Ragusa and Clara Troncon from the Milan ATLAS 
group.   


\end{document}

%% file: econfmacros.tex
\def\beq{\begin{equation}}
\def\eeq#1{\label{#1}\end{equation}}
\def\eeqn{\end{equation}}

\def\beqa{\begin{eqnarray}}
\def\eeqa#1{\label{#1}\end{eqnarray}}
\def\eeqan{\end{eqnarray}}

\let\bar=\overbar

\def\Dslash{\not{\hbox{\kern-4pt $D$}}}
\def\dslash{\not{\hbox{\kern-2pt $\del$}}}

\def\msb{{\bar{\ssstyle M \kern -1pt S}}}

\def\nequ{$\mbox{n}_{\mbox{\scriptsize eq}} \; \mbox{cm}^{-2}$}